%% file: paper.tex
\documentclass[12pt]{article}
\usepackage{epsfig}
\usepackage{amsmath}
\usepackage{hhline}
\usepackage{amssymb}
\usepackage{times}
\usepackage{cite}

\newlength{\dinwidth}
\newlength{\dinmargin}
\newcommand{\lsim}{\raisebox{-1.5mm}{$\:\stackrel{\textstyle{<}}{\textstyle{\sim}}\:$}
}

\def\GeV{\hbox{$\;\hbox{\rm GeV}$}}

\newcommand{\picob}{\mbox{{\rm ~pb}}}

\begin{document}
\begin{titlepage}

\noindent
DESY 01--094  \hfill  ISSN 0418-9833 \\
June 2001
\noindent

\vspace{2cm}

\begin{center}
\begin{Large}
\bf{A Search for Leptoquark Bosons
    in {\boldmath $e^- p$} Collisions at HERA \\ }

\vspace{2cm}

H1 Collaboration

\end{Large}
\end{center}

\vspace{2cm}

\begin{abstract}
\noindent

A search for scalar and vector leptoquarks coupling to first generation
fermions is performed in the H1 experiment at the $ep$ collider HERA.
The analysis uses $e^- p$ data collected in 1998 and
1999 at a centre-of-mass energy of $320 \GeV$, corresponding
to an integrated luminosity of $\sim 15 \picob^{-1}$.
No evidence for the direct production of such particles is found
in a data sample with a large transverse
momentum final state electron or with 
large missing transverse momentum,
and constraints on leptoquark models are established.
For a Yukawa coupling of electromagnetic strength
leptoquarks are excluded for masses up to $\sim 290 \GeV$.
This analysis complements the leptoquark searches performed previously
using data collected whilst HERA was operating with positrons
instead of electrons.
\end{abstract}

\vfill

\vfill

\begin{center}
 {To be submitted to {\em Phys. Lett. B}}
\end{center}

\end{titlepage}

\newpage

\include{h1auts_july01}
\newpage
\pagestyle{plain}


The $ep$ collider HERA offers the unique possibility to search for
resonant production of new particles which couple to
lepton-parton pairs.
Examples are leptoquarks (LQs), colour triplet bosons which appear naturally
in various unifying theories beyond the Standard Model (SM).
At HERA, leptoquarks could be singly produced by the fusion of the
initial state lepton of energy $27.5 \GeV$ with a quark from the
incoming proton of $920 \GeV$,
with masses up to the centre-of-mass energy $\sqrt{s_{ep}}$ of $320 \GeV$.

This analysis presents a search for LQs coupling to first
generation fermions using $e^- p$ data collected
in 1998 and 1999.
Collisions between {\it{electrons}} and protons provide a
high sensitivity to LQs with fermion number $F=2$ (i.e. LQs coupling to 
$e^-$ and a {\it{valence quark}})
while the production of such LQs is largely suppressed in
$e^+ p$ collisions where the interaction involves an 
{\it{antiquark}}\footnote{A fusion between an $e^+$ and a valence quark 
would lead to a LQ with $F=0$.}.
Thus this analysis complements the searches for LQs in $e^+ p$
data~\cite{H1LQ99,ZEUSLQ}.
This search considers the decays  ${\rm{LQ}} \rightarrow eq$ and
${\rm{LQ}} \rightarrow \nu q$ which lead to final states similar 
to those of deep-inelastic scattering (DIS) neutral current (NC)
and charged current (CC) interactions at very high
squared momentum transfer $Q^2$.
The integrated luminosity amounts to $15 \picob^{-1}$,
an increase in statistics by a factor of about~35 compared to
previous LQ searches~\cite{H1LQ94,ZEUSLQEMINUS} in $e^- p$ collisions.


The phenomenology of LQs at HERA was discussed in detail
in~\cite{H1LQ99}. At HERA, LQs can be resonantly produced in
the $s$-channel or exchanged in the $u$-channel between the
incoming lepton and a quark coming from the proton.
The amplitudes for both these processes interfere with those from DIS.
We shall consider here the mass domain where the resonant $s$-channel
contributions largely dominate the LQ signal cross-section.

In the $s$-channel, a LQ is produced at a mass $M =\sqrt{s_{ep} x}$
where $x$ is the momentum fraction of the proton carried by the 
interacting quark.
When the LQ decays into an electron and a quark, the mass 
is reconstructed from the measured kinematics of the scattered electron,
and is henceforth labelled $M_e$. Similarly when the LQ decays into a
neutrino and a quark, the mass is labelled $M_h$ as it is
reconstructed from the hadronic final state alone~\cite{H1LQ99}. 

The H1 detector components most relevant to this analysis are the liquid argon
calorimeter, which measures the positions and energies of
charged and neutral particles over
the polar angular range\footnote{The polar angle $\theta$ is defined with respect
to the incident proton momentum vector (the positive $z$ axis).}
$4^\circ<\theta<154^\circ$,
and the inner tracking detectors which measure
the angles and momenta of charged particles over the range
$7^\circ<\theta<165^\circ$. A full description of the detector can be
found in~\cite{h1det}.

\begin{figure}[tb]
  \begin{center}
  \begin{tabular}{cc}
     \hspace*{-0.2cm}
     \mbox{\epsfxsize=0.52\textwidth
        \epsffile{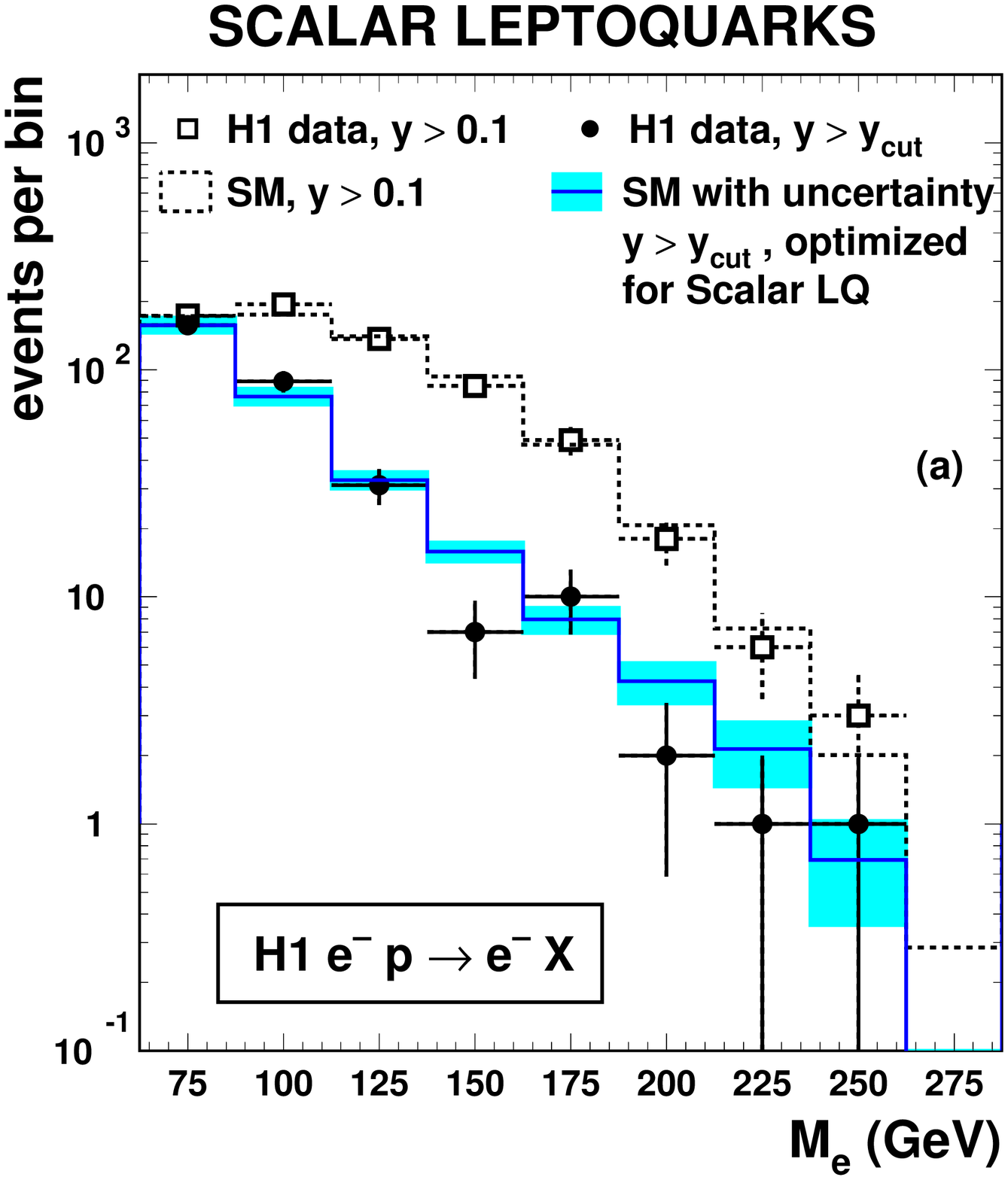}}
   &
     \hspace*{-0.8cm}\mbox{\epsfxsize=0.52\textwidth
       \epsffile{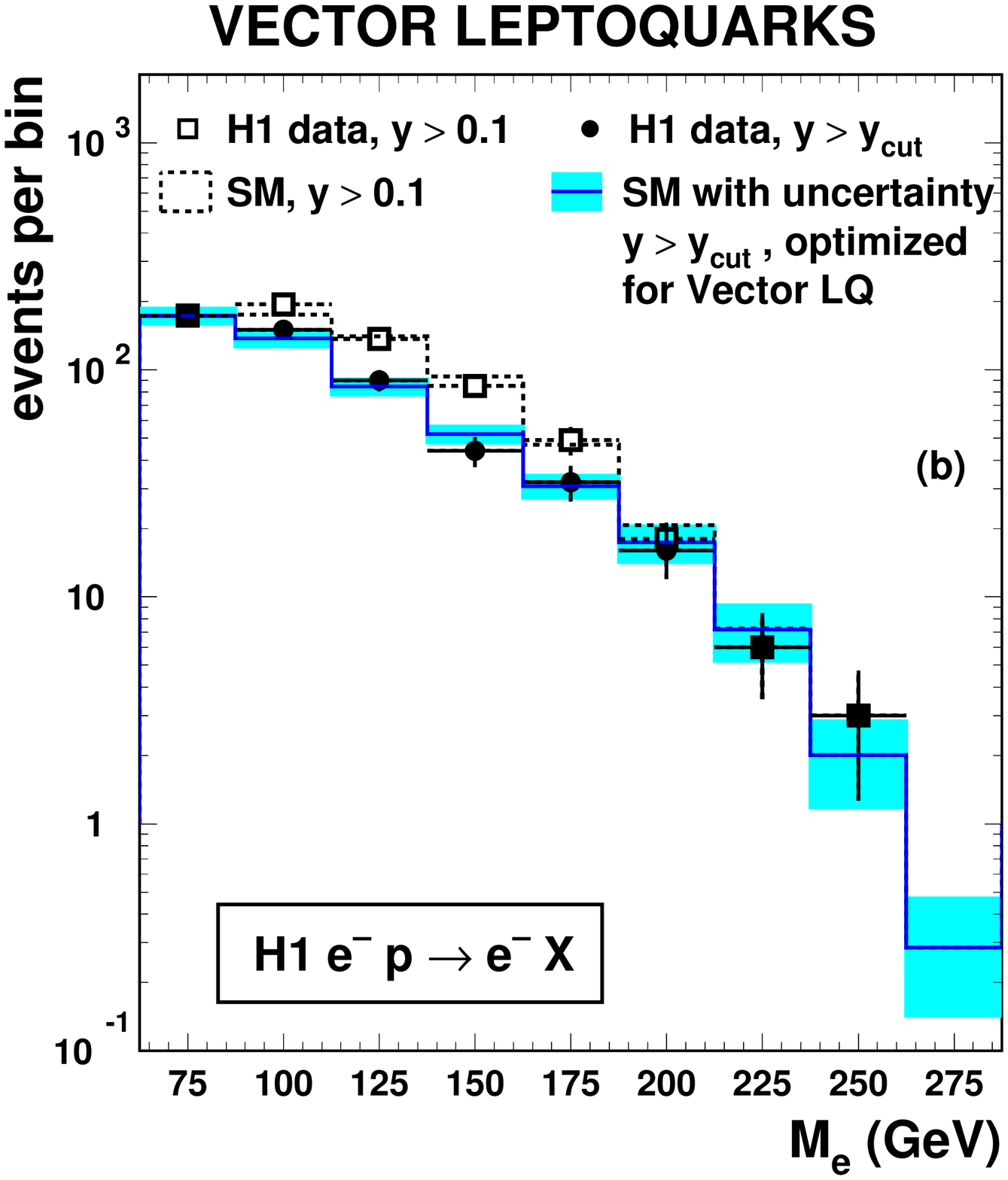}}
  \end{tabular}
  \end{center}
 \caption[]{ \label{fig:dndmnc}
 {\small Mass spectra of the events from the inclusive NC DIS selection
         for data (symbols) and DIS expectation (histograms).
         The data is shown before (open squares,
         dashed-line histogram) and after (filled dots, full-line histogram)
         a $y$ cut designed to maximize the significance of 
         (a) a scalar and (b) a vector leptoquark (LQ) signal.
         The grey boxes indicate the $\pm 1 \sigma$ uncertainty  
         due to the systematic errors on the NC DIS
         expectation. }}
\end{figure}

This search relies essentially on
inclusive NC and CC DIS selections.
The selection of NC-like events is identical to that presented
in~\cite{H1LQ99}.
It requires an identified electron with transverse energy above
$15 \GeV$ and considers the kinematic domain defined by
$Q^2 > 2500 \GeV^2$ and $0.1 < y < 0.9$, where
$y=Q^2/M^2$.
The inelasticity variable $y$ is related to 
the polar angle $\theta^\ast$ of the lepton
in the centre-of-mass frame of the hard subprocess
by $y =\frac{1}{2}(1+\cos\theta^\ast)$.
Since the angular distribution of the electron coming from the
decay of a scalar (vector) resonance is markedly (slightly)
different from that of the scattered lepton in NC DIS~\cite{H1LQ99}, a mass   
dependent cut $y>y_{\rm cut}$ allows   
the signal significance to be optimized.
The measured mass spectrum is compared in Fig.~\ref{fig:dndmnc}
with the NC SM prediction, obtained using a Monte-Carlo
calculation~\cite{DJANGO} and the MRST parametrization~\cite{mrst} 
for the parton densities.
The distributions are shown before and after applying the
mass dependent lower $y$ cut
designed to maximize the significance of a scalar (Fig.~\ref{fig:dndmnc}a)
or vector (Fig.~\ref{fig:dndmnc}b) LQ.
For scalar (vector) LQs, $y_{\rm cut}$ continuously decreases
from $\sim 0.45$ ($\sim 0.25$) at $100 \GeV$
to $\sim 0.35$ ($\sim 0.15$) at 200 GeV, reaching 0.1 (0.1)
at 290 GeV.
In the mass range $M_e>62.5$\,GeV and after applying
the $y$ cut optimized for scalar (vector) LQ searches,
298 (514) events are 
observed in good agreement with the SM expectation of $297 \pm 22$   
($504 \pm 38$) events.
%
%
%
%
%
\begin{figure}[h]
   \begin{center}
     \epsfxsize=0.52\textwidth
      \epsffile{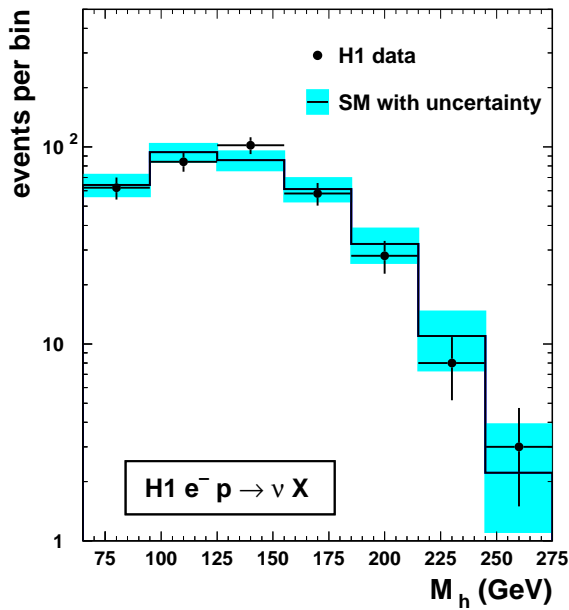}
      \caption
      { \small  \label{fig:dndmcc}
       Mass spectra  of the events from the inclusive CC DIS selection
         for data (symbols) and DIS expectation (histogram).
         The grey boxes indicate the $\pm 1 \sigma$ uncertainty 
         due to the systematic errors on the CC DIS
         expectation. }
 \end{center}
\end{figure}
%

The selection of CC-like events follows closely that
presented in~\cite{H1EMINUS}.
In addition, a missing transverse momentum exceeding $25 \GeV$
and $Q^2 > 2500 \GeV^2$ are required.
The domain at high $y$ where
the resolution on the mass $M_h$ degrades is removed by requiring $y<0.9$. 
For $M_h > 65 \GeV$, 345 events are observed, in good agreement with
the CC SM expectation of $350 \pm 28$ events.
The observed and expected mass spectra are shown
in Fig.~\ref{fig:dndmcc}.

No evidence for LQ production is observed in either data sample.
Hence the data are 
used to set constraints on LQs which couple
to first generation fermions.
We use the numbers of observed and expected events within a
variable mass bin, adapted to the experimental mass distribution
for a given true LQ mass $M_{\rm LQ}$, and
which slides over the accessible mass range.
As an example, candidate events with $M_e$
within the interval from $187 \GeV$ to $206 \GeV$ 
are used to constrain
a $200 \GeV$ LQ decaying into electrons.
For LQs decaying into $\nu q$, the mass window is enlarged
(to about $40$~GeV for a 200~GeV LQ) to account for the
mass resolution when relying on the hadronic final state.
The final signal efficiencies, including the mass bin
requirement, vary with the LQ mass between $35 \%$ ($20 \%$) and
$52 \%$ ($45 \%$) for scalar (vector) LQs decaying into $eq$,
and between $20 \%$ and $52 \%$ for LQs decaying into $\nu  q$.

Assuming Poisson distributions for the SM background expectations and
for the signal, an upper limit on the number of events coming
from LQ production is obtained using a standard Bayesian
prescription. This limit on the number of signal events
is then translated into an upper bound on the LQ cross-section,
which in turn leads to constraints on LQ models.
The signal cross-section is obtained from the leading-order 
LQ amplitudes given in~\cite{BRW},
corrected by multiplicative $K$-factors~\cite{LQNLO} to account
for next-to-leading order QCD corrections.
These corrections can enhance the LQ cross-section by ${\cal{O}}(10 \%)$.

The procedure which folds in the
statistical and systematic errors is described in detail
in~\cite{H1LQ94}.
The main source of experimental systematic error is the uncertainty
on the electromagnetic energy scale (between $0.7 \%$ and
$3 \%$) for the NC analysis, and the
uncertainty on the hadronic energy scale ($2 \%$)
for the CC analysis.
Furthermore, an error of $\pm 7 \%$ on the DIS expectations is attributed
to the limited knowledge of proton structure. An additional systematic 
error arises from the theoretical uncertainty on the signal cross-section,
originating mainly from the uncertainties on the parton densities.
This uncertainty is $7 \%$ for LQs coupling to $e^- u$, and varies 
between $7 \%$ 
at low LQ masses up to $50 \%$ around 290\,GeV for LQs coupling to $e^- d$.
Moreover, choosing alternatively $Q^2$ or the square of the transverse momentum
of the final state lepton instead of $M_{\rm LQ}^2$ as the hard scale at which
the parton distributions are estimated yields an additional uncertainty of
$\pm 7 \%$ on the signal cross-section.

\begin{figure}[htb]
   \begin{center}
     \epsfxsize=0.9\textwidth
     \epsffile{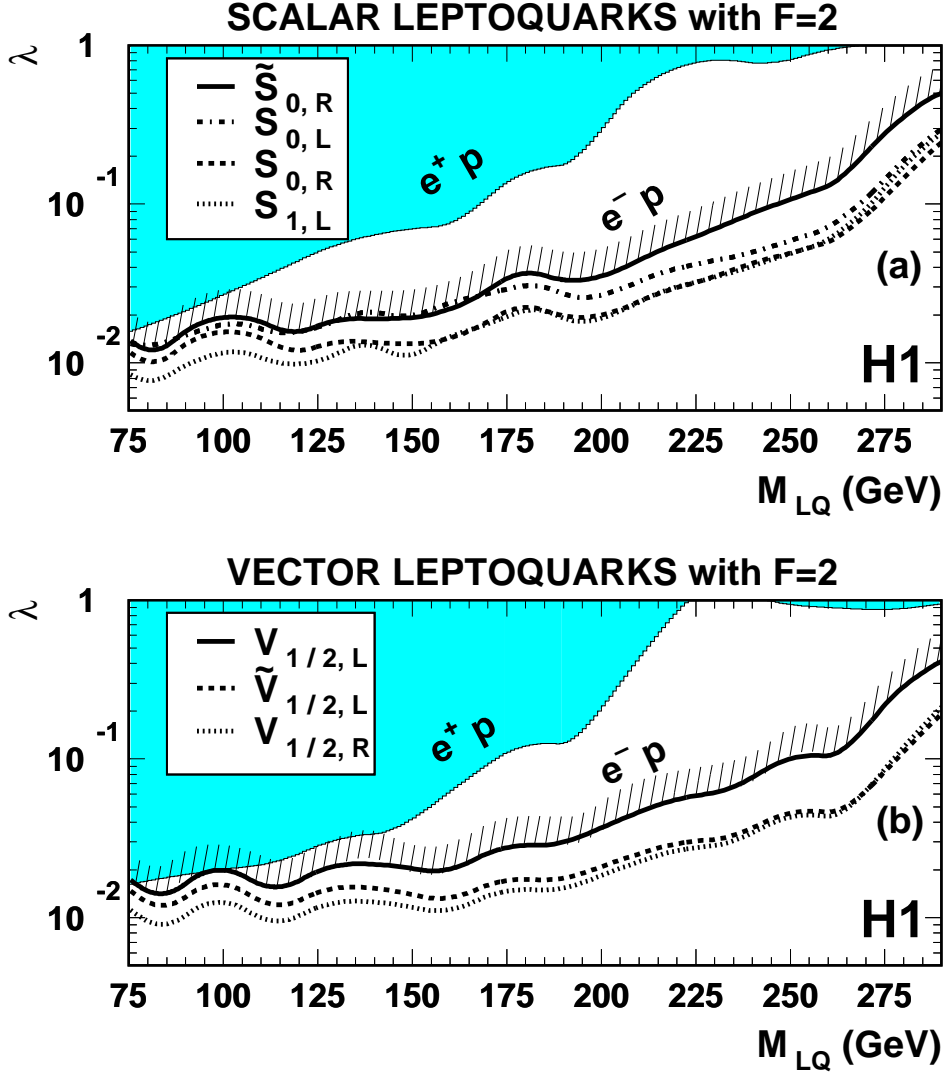}
      \caption
      { \small  \label{fig:brw}
                Exclusion limits at $95 \%$ CL on the Yukawa
                coupling $\lambda$ as a function of the mass
                of (a) scalar and (b) vector leptoquarks (LQs) with
                fermion number $F=2$ described by the BRW model.
                Domains above the curves are excluded by this 
                analysis of the $e^- p$ data.
                The shaded area on each plot indicates the excluded region
                obtained from the 
                $e^+ p$ data~\cite{H1LQ99}, less suited 
                for constraining $F=2$ LQs.}
 \end{center}
\end{figure}

The phenomenological model proposed by
Buchm\"uller, R\"uckl and Wyler (BRW)~\cite{BRW}
describes 14 LQs.
We focus here on the 7 LQs with fermion number $F=2$ since
those with $F=0$ are better constrained using $e^+ p$ data~\cite{H1LQ99}.
In the BRW model the branching ratios $\beta_e$ ($\beta_{\nu}$)
for the LQ decays
into $e q$ ($\nu q$) are fixed and equal
to 1 or 0.5 (0 or 0.5) depending on the LQ quantum numbers.
The upper limits on the Yukawa coupling $\lambda$ at the 
$e \, q \, \rm{LQ}$ vertex
obtained at $95 \%$ confidence level (CL) are
shown as a function of the LQ mass in Figs.~\ref{fig:brw}a
and b, for scalar and vector LQs
respectively. The nomenclature of~\cite{LQNAME}
is used to label the various scalar $S_{I,L}$ 
($\tilde{S}^{\mbox{\tiny \hspace{-3mm}\raisebox{1.5mm}{(}\hspace{2mm}\raisebox{1.5mm}{)}}}_{I,R}$)
or vector $\tilde{V}^{\mbox{\tiny \hspace{-3mm}\raisebox{1.5mm}{(}\hspace{2mm}\raisebox{1.5mm}{)}}}_{I,L}$ ($V_{I,R}$) LQ types of weak 
isospin $I$, which
couple to a left-handed (right-handed) electron. The tilde is used to 
distinguish LQs which differ only by their hypercharge.
%
%
%
%
For LQs decaying with an equal branching ratio into $e q$
and $\nu q$,
both the NC and CC channels were combined 
in the derivation of the limits. 
However, the CC channel offers much less sensitivity 
to the signal than the NC
channel, and thus only marginally contributes to the resulting bounds.
%
This is due to the fact that the mass windows are larger, and that no
discriminating angular cut is applied in the CC channel.
Both effects yield a much larger SM background in the CC
channel than in the NC case\footnote{
  This is different from the $e^+ p$ case, where the CC channel
  significantly improves the sensitivity on the LQ production
  cross-section~\cite{H1LQ99} due to the smaller CC DIS cross-section.}.
For a Yukawa coupling of electromagnetic strength $\alpha_{\rm em}$
($\lambda = \sqrt{4\pi\alpha_{\rm em}} \simeq 0.3$)
this analysis rules out LQ masses below $275$ to $290 \GeV$
depending on the LQ type, at $95 \%$ CL.
These are the most stringent direct bounds on LQs with $F=2$.

\begin{figure}[htb]
   \begin{center}
    \begin{tabular}{c}
     \mbox{\epsfxsize=0.8\textwidth
      \epsffile{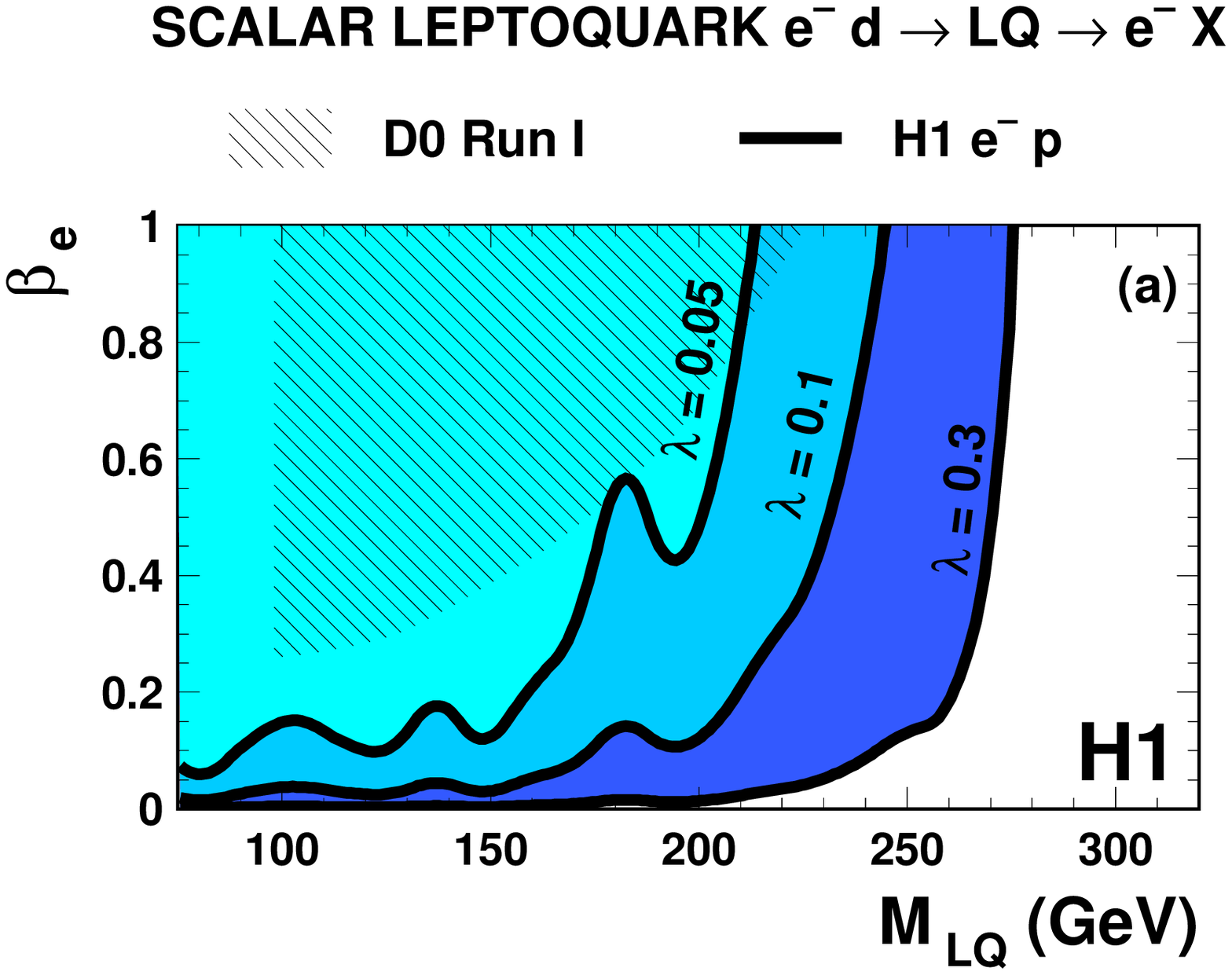}} \\
     {\raisebox{-70pt}{\mbox{\epsfxsize=0.8\textwidth
     \epsffile{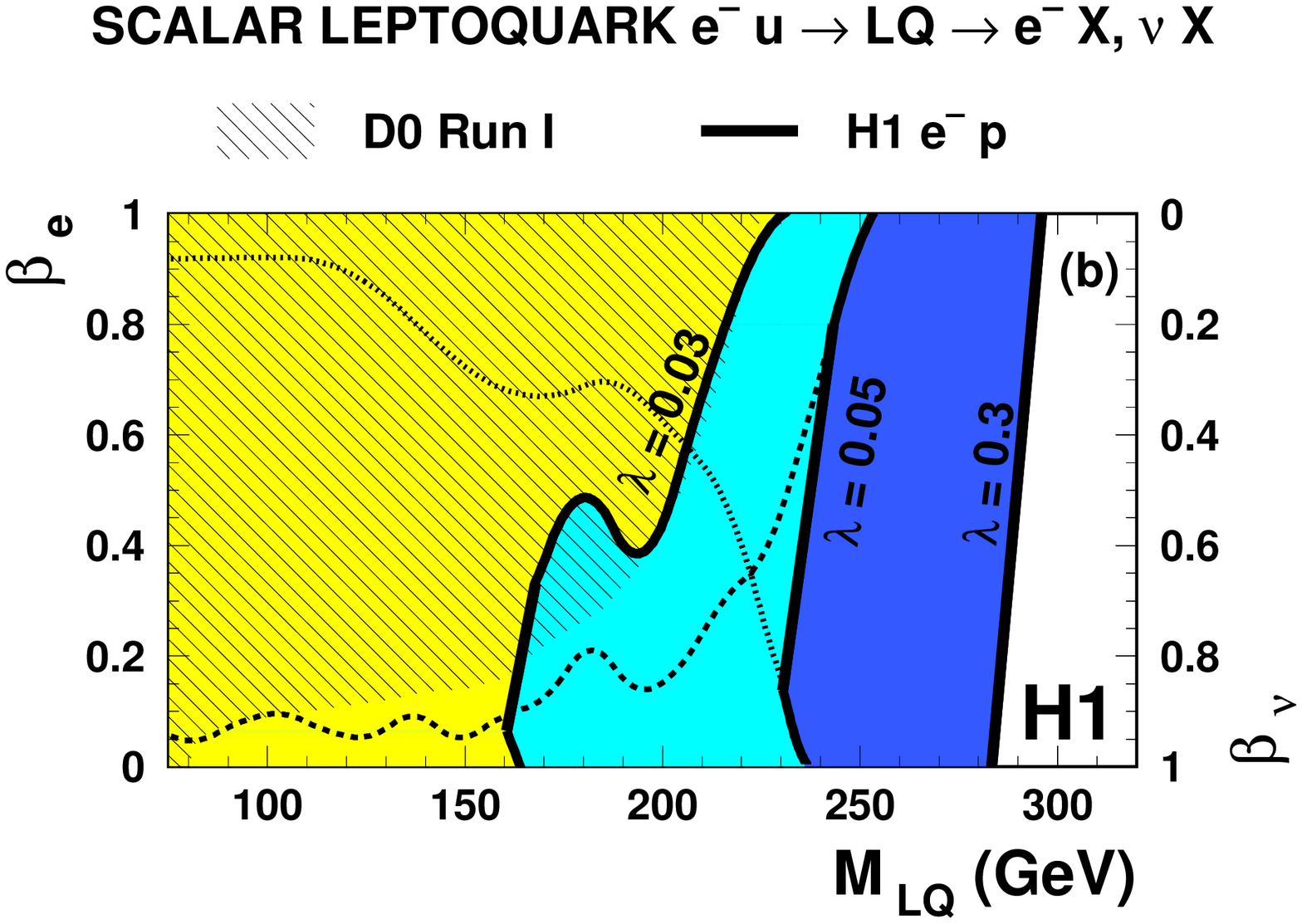}}
     }}
    \end{tabular}
      \caption
      { \small  \label{fig:beta}
         (a) Mass dependent exclusion limits at $95 \%$ CL on the branching
         ratio $\beta_e$ of a scalar leptoquark (LQ) which couples to $e^- d$ 
         (with the quantum numbers of the $\tilde{S}_{0,R}$).
       (b) Domains ruled out by the combination of the NC and CC analyses,
       for a scalar LQ which couples to $e^- u$ (with the quantum
       numbers of the $S_{0,L}$)  and
       decaying only into $e q$ and $\nu q$
       for three example values
       of the Yukawa coupling $\lambda$.
       The regions on the left of the full curves
       are excluded at $95 \%$ CL.
       For $\lambda=0.05$, the part of the $\beta_e$-$M_{{\rm{LQ}}}$
       ($\beta_{\nu}$-$M_{{\rm{LQ}}}$) plane on the left of the dashed
       (dotted) curve is excluded by the NC (CC) analysis.
       The branching ratios $\beta_e$ and $\beta_{\nu}$ are shown
       on the left and right axes respectively.
       In (a) and (b), the hatched region represents the domain
       excluded by the D$0$ experiment~\cite{d0}.}
 \end{center}
\end{figure}

Beyond the BRW ansatz, generic LQ models can also be considered, where
other LQ decay modes are allowed such that the branching ratios $\beta_e$ 
and $\beta_\nu$ are free parameters.
The LQ production cross-section does not depend on
the total LQ width $\Gamma$ as long as $\Gamma$ is not too large.
Hence the signal cross-section observable in e.g. the
NC channel
depends only on the Yukawa coupling and on the
branching ratio $\beta_e$, and  mass dependent
constraints on $\beta_e$ can be set for a given
value of $\lambda$.
For a scalar LQ with $M_{\rm{LQ}} = 295 \GeV$ and $\lambda=0.3$, this approach 
holds as long as $\Gamma \lsim 2 \GeV$, such that the LQ
total width does not exceed about four times its partial decay width
into $e q$.
For a scalar LQ possessing the quantum numbers of the
$\tilde{S}_{0,R}$, which couples
to $e^- d$ and thus cannot decay into $\nu q$, 
Fig.~\ref{fig:beta}a shows the excluded part of the
$\beta_e$-$M_{{\rm{LQ}}}$ plane for three values of  the Yukawa coupling.
The domain excluded by the 
D$0$ experiment at the Tevatron~\cite{d0} is also shown.
%
%
%
For a scalar LQ coupling to $e^- u$ (possessing the
quantum numbers of the $S_{0,L}$) and for $\lambda = 0.05$, 
the domain of the $\beta_e$-$M_{\rm LQ}$ 
($\beta_\nu$-$M_{\rm LQ}$) plane excluded by the NC (CC) analysis
is shown in Fig.~\ref{fig:beta}b.
If the LQ decays into $e q$ or $\nu  q$
only\footnote{It should be noted that $\beta_e + \beta_{\nu} = 1$ does
   not imply $\beta_e = \beta_{\nu}$ even when 
   invariance under $SU(2)_L$ transformations is required.
   For example, when LQs belonging to a given isospin multiplet are not
   mass eigenstates, their mixing usually leads to different branching
   ratios in both channels for the physical LQ states. },
the combination
of both channels rules out the part of the plane on the left of
the middle full curve, for $\lambda = 0.05$. The resulting
combined bound is largely independent of the individual values
of $\beta_e$ and $\beta_{\nu}$.
Combined bounds are also shown for $\lambda=0.03$ and $\lambda=0.3$,
for the same LQ type. 
For $\lambda$ greater than $\sim 0.03$, these bounds extend considerably beyond the
region excluded by the D$0$ experiment~\cite{d0}.

To summarize, a search for leptoquarks with fermion number $F=2$
has been performed using the $e^- p$ data collected by H1 in 1998 and 1999.
No signal has been observed and constraints on such LQs have been
set, which extend beyond the domains excluded by other experiments.
For a Yukawa coupling of electromagnetic strength, LQ masses
up to 290 GeV can be ruled out.
This represents the most stringent direct bound on $F=2$ leptoquarks.

\section*{Acknowledgements}
We are grateful to the HERA machine group whose outstanding
efforts have made and continue to make this experiment possible.
We thank
the engineers and technicians for their work in constructing and now
maintaining the H1 detector, our funding agencies for
financial support, the
DESY technical staff for continual assistance
and the DESY directorate for the
hospitality which they extend to the non DESY
members of the collaboration.

\end{document}

%% file: h1auts_july01.tex

\noindent
C.~Adloff$^{33}$,              
V.~Andreev$^{24}$,             
B.~Andrieu$^{27}$,             
T.~Anthonis$^{4}$,             
V.~Arkadov$^{35}$,             
A.~Astvatsatourov$^{35}$,      
A.~Babaev$^{23}$,              
J.~B\"ahr$^{35}$,              
P.~Baranov$^{24}$,             
E.~Barrelet$^{28}$
% ...

\bigskip{\noindent \it
% ...
$ ^{4}$ Inter-University Institute for High Energies ULB-VUB, Brussels;
          Universitaire Instelling Antwerpen, Wilrijk; Belgium$^{ c}$ \\
% ...
$ ^{23}$ Institute for Theoretical and Experimental Physics,
          Moscow, Russia$^{ l}$ \\
 $ ^{24}$ Lebedev Physical Institute, Moscow, Russia$^{ e,h}$ \\
% ...
$ ^{27}$ LPNHE, Ecole Polytechnique, IN2P3-CNRS, Palaiseau, France \\
 $ ^{28}$ LPNHE, Universit\'{e}s Paris VI and VII, IN2P3-CNRS,
          Paris, France \\
% ...
$ ^{33}$ Fachbereich Physik, Bergische Universit\"at Gesamthochschule
          Wuppertal, Wuppertal, Germany$^{ a}$ \\
% ...
$ ^{35}$ DESY, Zeuthen, Germany$^{ a}$ \\
% ...
$ ^a$ Supported by the Bundesministerium f\"ur Bildung, Wissenschaft,
      Forschung und Technologie, FRG,
      under contract numbers 7AC17P, 7AC47P, 7DO55P, 7HH17I, 7HH27P,
      7HD17P, 7HD27P, 7KI17I, 6MP17I and 7WT87P \\
% ...
$ ^c$ Supported by FNRS-NFWO, IISN-IIKW \\
% ...
$ ^e$ Supported by the Deutsche Forschungsgemeinschaft \\
% ...
$ ^h$ Supported by Russian Foundation for Basic Research
      grant no. 96-02-00019 \\
% ...
$ ^l$ Partially Supported by Russian Foundation
      for Basic Research, grant    no. 00-15-96584 \\
}